## VSX J003909.7+611233: a new Slowly Pulsating B star (SPB) in Cassiopeia?

David Boyd, Christopher Lloyd, Pierre dePonthiere, Mack Julian, Robert Koff, Tom Krajci, Jeremy Shears, Bart Staels

**Abstract**

We report the discovery of a new 13$^{th}$ magnitude variable in Cassiopeia close to the variable KP Cas. Analysis of 6 days of intensive photometry shows a regular, near sinusoidal modulation with an amplitude of 0.024 magnitudes and a period of 0.43815(31) d. Although its colour indicates a spectral type around F0 the star probably suffers up to 2 – 2.5 magnitudes of extinction so could be an A- or B-type star. Given the period, the low amplitude, the shape of the light curve and the probable spectral type we consider it most likely to be a slowly pulsating B-type (SPB) star. The variable has been registered in the International Variable Star Index with the identifier VSX J003909.7+611233.

**Discovery**

During observation of the 2008 October superoutburst of KP Cassiopeia [1], we observed variability in a nearby 13$^{th}$ magnitude star which had initially been chosen as a comparison star. We measured its position using Astrometrica [2] and USNO-B1.0 as 00h 39m 09.796+/-0.015s +61º 12' 33.54+/-0.19'' (J2000). Inspection of Vizier [3] reveals catalogue entries at this position with the following identifiers: USNO-A2.0 1500-00686865; USNO-B1.0 1512-0023462, 2MASS J00390981+6112331 and GSC2.3 NALV008506. No previous record of its variability was found in GCVS [4] or VSX [5] and the star is not listed in Simbad [6]. The 2MASS magnitudes are J = 12.61, H = 12.49, K = 12.43 [7]. We therefore believe the variability of this star is a new discovery and the variable has been registered in VSX with the identifier VSX J003909.7+611233. Figure 1 shows the field around KP Cas with the new variable marked.

**Analysis**

A log of 19 observing runs contributing to this analysis is given in Table 1. Details of the equipment used are given in [1] and include a mixture of filtered and unfiltered systems. Images were dark-subtracted and flat-fielded and a measurement of the magnitude of the new variable was obtained from each of 8110 images using the same comparison stars used for KP Cas [1]. Heliocentric corrections were applied to all times of observation. The magnitude calibration of each observer's data was slightly different because of differences in the spectral responses of their equipment. A preliminary analysis of our data showed the amplitude of variation was 0.02-0.03 magnitude so it was clear that careful alignment of all observers' data in magnitude would be necessary to obtain an accurate measurement of the period and amplitude.

All light curves were inspected to locate the positions of maxima and minima. The time and magnitude at each extremum were obtained by a weighted second order polynomial fit to the light curve around the extremum. From these fitted maximum and minimum magnitudes, adjustments were applied to each run to bring the magnitudes of all runs into mutual alignment. The average size of these adjustments was 0.007 magnitudes. The resulting combined light curve over the 6 day interval October 27 to November 1 is shown in Figure 2. The Discrete Fourier Transform (DFT) power spectrum of the data is shown in Figure 3 and this reveals a clear single periodic variation at f=2.27 c/d or P=0.44 days, with a full amplitude of 0.02 magnitudes.

The period was refined in two stages. Firstly, cycle numbers were assigned to each maximum and minimum by inspection of the light curve. A weighted linear fit to these times gave the following linear ephemeris for the times of maximum light

$$HJD(max) = 2454767.4648(45) + 0.43700(62) * E \quad (1)$$

and the observed minus calculated (O-C) times with respect to this ephemeris are shown in Figure 4. This gave an improved period estimate of 0.4370 days.

The internal scatter in the data from each observer ranged from 0.005 to 0.022 magnitudes according to the equipment used and the conditions prevailing at their observing site. This scatter was very consistent for a single observer from night to night. The datasets with larger scatter were first smoothed by averaging adjacent data points to make the internal errors consistent across the datasets. The combined datasets were then fitted with a Fourier series at the above period and also at twice this period, which would be appropriate for an eclipsing binary light curve, and in both cases small magnitude offsets were allowed for each dataset to find the best fit. The corresponding ephemerides are

$$HJD(max) = 2454770.5218(36) + 0.43815(31) * E \quad (2)$$
$$HJD(min1) = 2454770.7445(46) + 0.87602(61) * E \quad (3)$$

and the phase diagrams for these two solutions are shown in Figures 5 and 6. The data plotted in these figures are 5-point medians and the mean amplitudes are 0.024 and 0.021 magnitudes respectively. The light curve is basically sinusoidal but the minimum is rather pointed and the rise to maximum is distorted. Although these phased signals appear quite well defined, their amplitude is small and caution needs to be exercised as the photometry is a combination of V filtered and unfiltered data and the amplitude of the variable may change with wavelength.

Very little is known about the star although there are colours from 2MASS [7] which give J-H = 0.12 and H-K = 0.06. Our measurements of the B-V and V-R colour indices give 0.39(1) and 0.23(1) respectively. All these values suggest a spectral type near F0 if the star is unreddened, however, to give the observed V magnitude a main sequence star of this spectral type would lie at a distance of some 1500 parsecs. As the star lies within two degrees of the galactic plane there will be some reddening so the spectral type will be earlier than suggested by the colours. Measurements of the reddening in this direction suggest $E_{(B-V)} \sim 0.4$ for NGC 189 [8] which lies at 750 pc and $E_{(B-V)} \sim 0.6$ for the Cas OB4 association which lies at about 3 kpc [9]. With B-V = 0.39 this amount of reddening means that the variable must be at least as early as spectral type A0 and might be as early as B2. The V magnitude, colours and reddening are most consistent with the variable lying at some distance, probably 2 – 3 kpc, which places it in the Perseus-Cassiopeia spiral arm. The 2MASS colours limit the visual extinction to $A_V < 2.5$ magnitudes so an earlier, more luminous star, say B0 would have to lie at several kpc and in this direction that seems unlikely.

Given the range of possible spectral types this star could, on the face of it, be almost any of the low-amplitude variable stars on the upper main sequence (see Degroote et al. for recent references [10]). If the star is very early B-type then it may be a beta (β) Cephei variable but at 0.44 days the period would be extremely long [11] and from the previous discussion the luminosity would be at the upper limit of acceptability. The Slowly Pulsating B-type (SPB) stars occupy the range from B2 – B9 and although these typically have periods of 1 – 4 days this variable does lie within the short-period limit [12]. If the star is a chemically peculiar Bp or Ap star then low amplitude variations could be produced by rotation but again the periods of these variables are typically a few days [13]. Cooler pulsating variables like the gamma (γ) Doradus [14] and delta (δ) Scuti [15] variables are generally F-type stars and can probably be excluded on colour and luminosity grounds. The δ Scuti variables also have much shorter periods. Finally, the variable may be an eclipsing binary or ellipsoidal variable seen at low inclination. Given the period and likely spectral type the circumstances would have to be very specific to produce such a low-amplitude variation.

On the available evidence, there is little to choose between the photometric solutions with one or two minima. Since the two halves of the longer period are not significantly different in shape or amplitude this solution does not recommend itself, but at the same time it cannot be excluded. If it were an eclipsing binary it is not clear what could cause similar distortions in both halves of the light curve. Given the shape of the light curve the most likely interpretation is a pulsating variable and the probable spectral type suggests an SPB star. These variables frequently show multiple periods but no other significant periods have been found in the data analysed here. Any other periods must have amplitudes below 0.005 magnitudes although it is conceivable that larger variations could show up in longer runs of data

**Conclusion**

A new variable has been found in the field of KP Cas with an amplitude of 0.024 magnitude and a period of 0.43815(31) days. It is probably reddened by about half a magnitude and most likely lies in or near the Perseus-Cassiopeia spiral arm. The shape of the light curve and the probable spectral type suggests it is an SPB star.

**Acknowledgements**

We acknowledge with thanks use of the Simbad and Vizier services operated by CDS Strasbourg. We are also grateful for helpful comments from the referee and we are indebted to John Greaves whose comments have improved the paper.

**Addresses**

DB: 5 Silver Lane, West Challow, Wantage, Oxon, OX12 9TX, UK [drsboyd@dsl.pipex.com]
CL: Department of Physics and Astronomy, Open University, Milton Keynes, MK7 6AA, UK [c.lloyd@open.ac.uk]
PdeP: 15 rue Pre Mathy, 5170 Lesve-Profondeville, Belgium [pierredeponthiere@gmail.com]
MJ: 4587 Rockaway Loop, Rio Rancho, NM 871224, USA [mack-julian@cableone.net]
RK: CBA Colorado, 980 Antelope Drive West, Bennett, CO 80102, USA [bob@antelopehillsobservatory.org]
TK: CBA New Mexico, PO Box 1351 Cloudcroft, New Mexico 88317, USA [tom_krajci@tularosa.net]
JS: “Pemberton”, School Lane, Bunbury, Tarporley, Cheshire, CW6 9NR, UK [bunburyobservatory@hotmail.com]
BS: CBA Flanders (Patrick Mergan Observatory), Koningshofbaan 51, B-9308 Hofstade, Belgium [staels.bart.bvba@pandora.be]

| Start time (JD) | Duration (hrs) | Filter | Observer |
|---|---|---|---|
| 2454767.26039 | 7.70 | V | Boyd |
| 2454767.36072 | 4.10 | C | Shears |
| 2454767.61938 | 8.13 | V | Julian |
| 2454768.24024 | 7.14 | C | Shears |
| 2454768.26340 | 6.40 | C | Staels |
| 2454768.57255 | 10.48 | V | Koff |
| 2454769.22919 | 4.17 | C | dePonthiere |
| 2454769.23649 | 3.67 | C | Staels |
| 2454769.54403 | 10.60 | V | Koff |
| 2454769.60627 | 8.44 | V | Julian |
| 2454770.27623 | 8.24 | C | Boyd |
| 2454770.28947 | 5.77 | C | Shears |
| 2454770.53278 | 11.01 | V | Koff |
| 2454770.60228 | 8.54 | V | Julian |
| 2454771.23582 | 7.97 | C | Boyd |
| 2454771.26962 | 7.00 | C | Shears |
| 2454771.65566 | 7.56 | C | Krajci |
| 2454772.23536 | 1.56 | C | Shears |
| 2454772.42193 | 7.35 | C | dePonthiere |

Table 1. Log of observing runs.

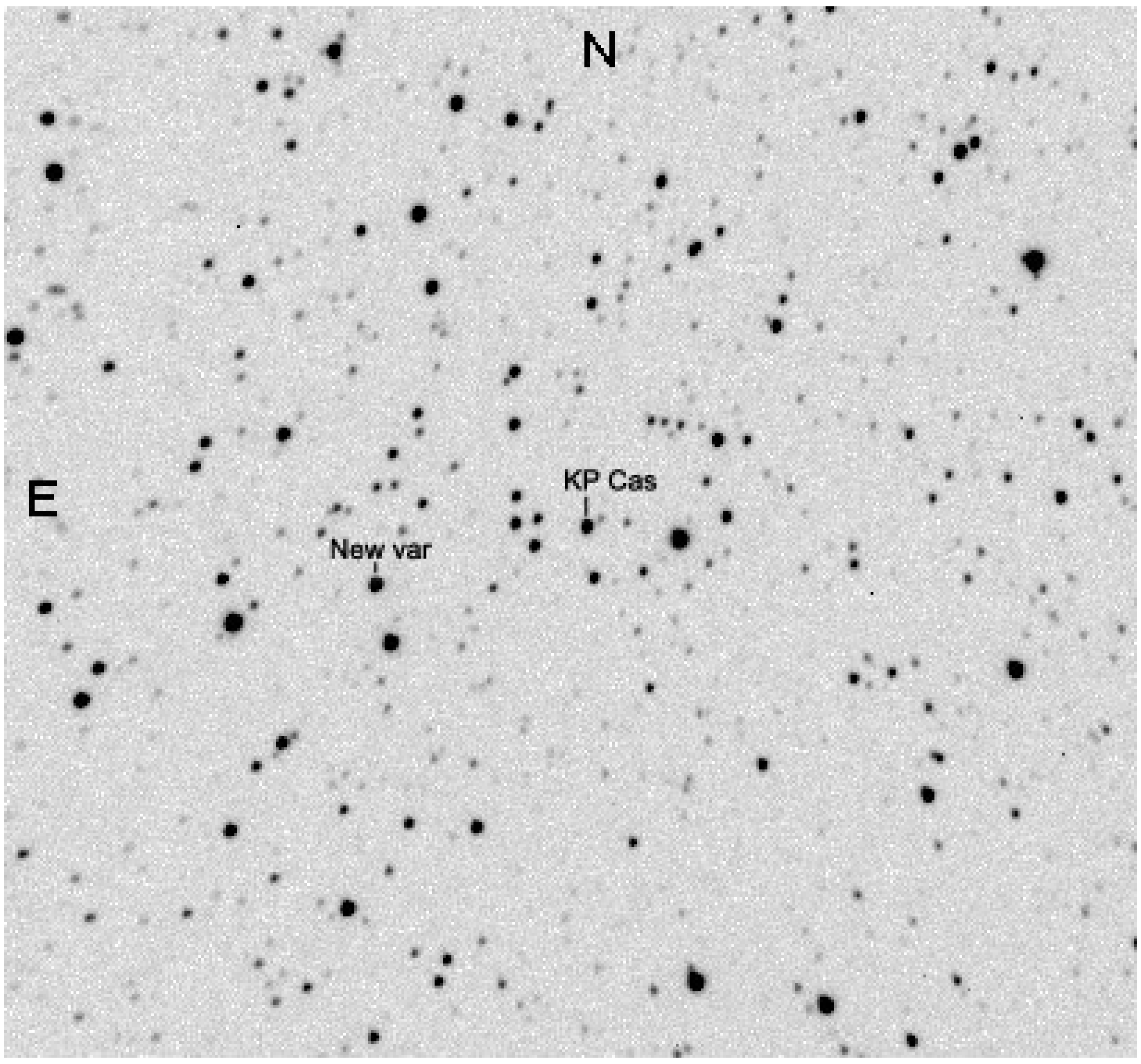

Figure 1. Location of the new variable close to KP Cas. V filtered image taken on 2008 October 27. Field 10' x 10' (*Boyd*).

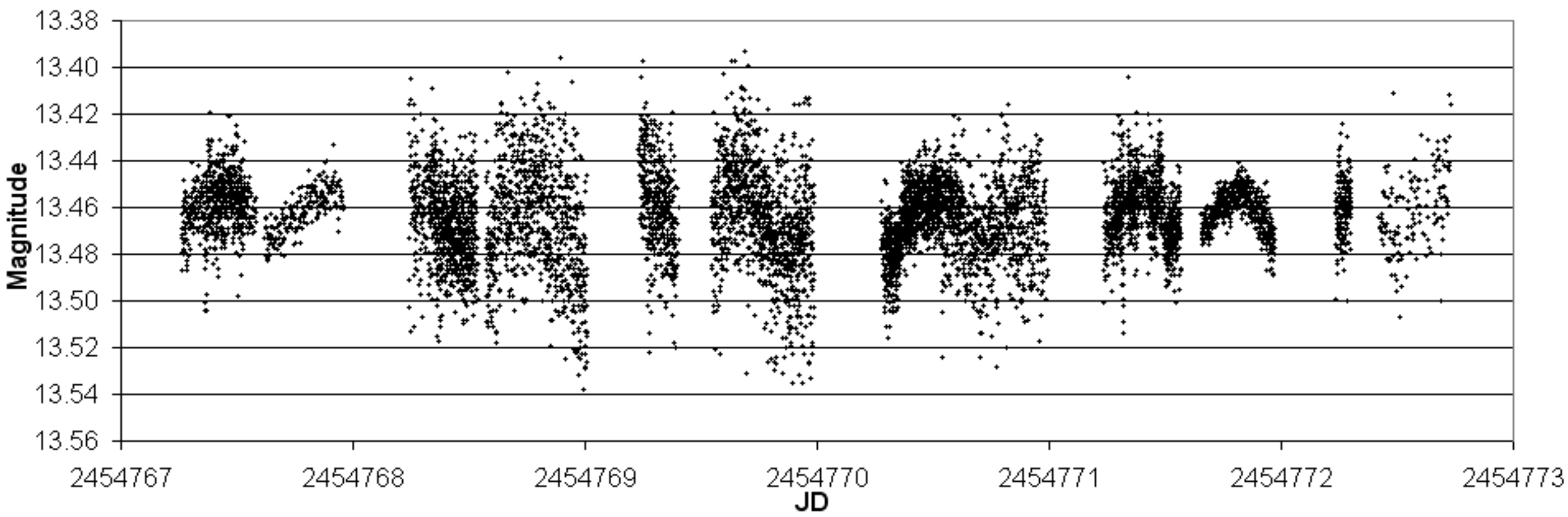

Figure 2. Combined light curve.

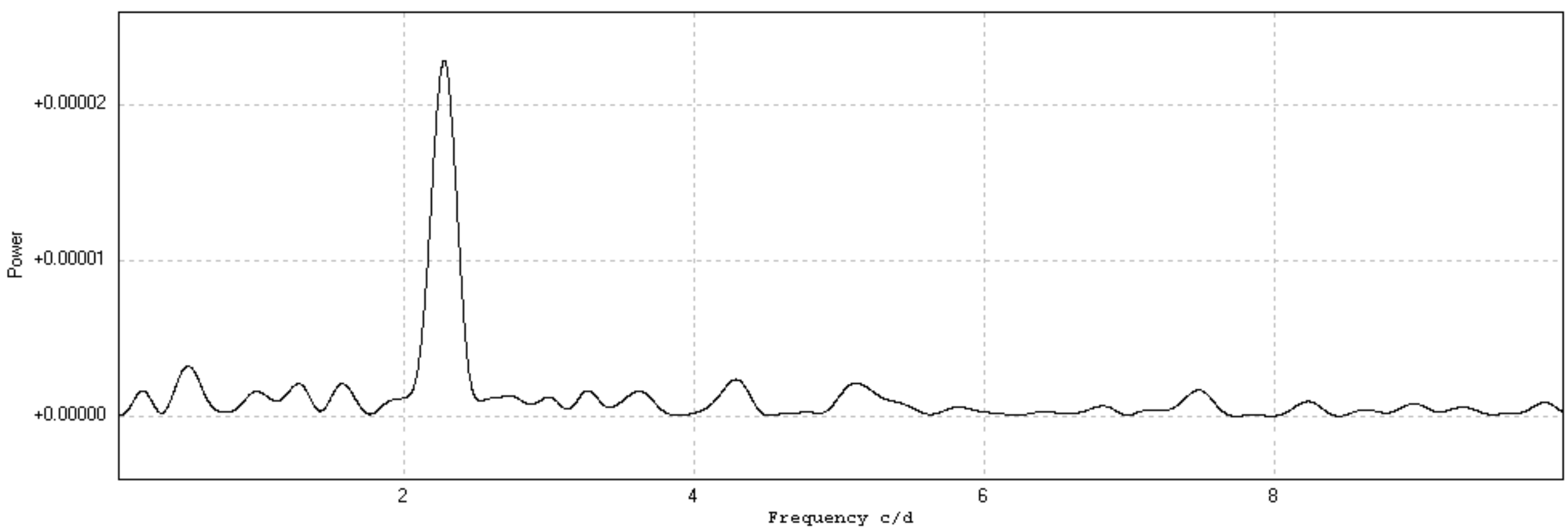

Figure 3. Discrete Fourier Transform power spectrum.

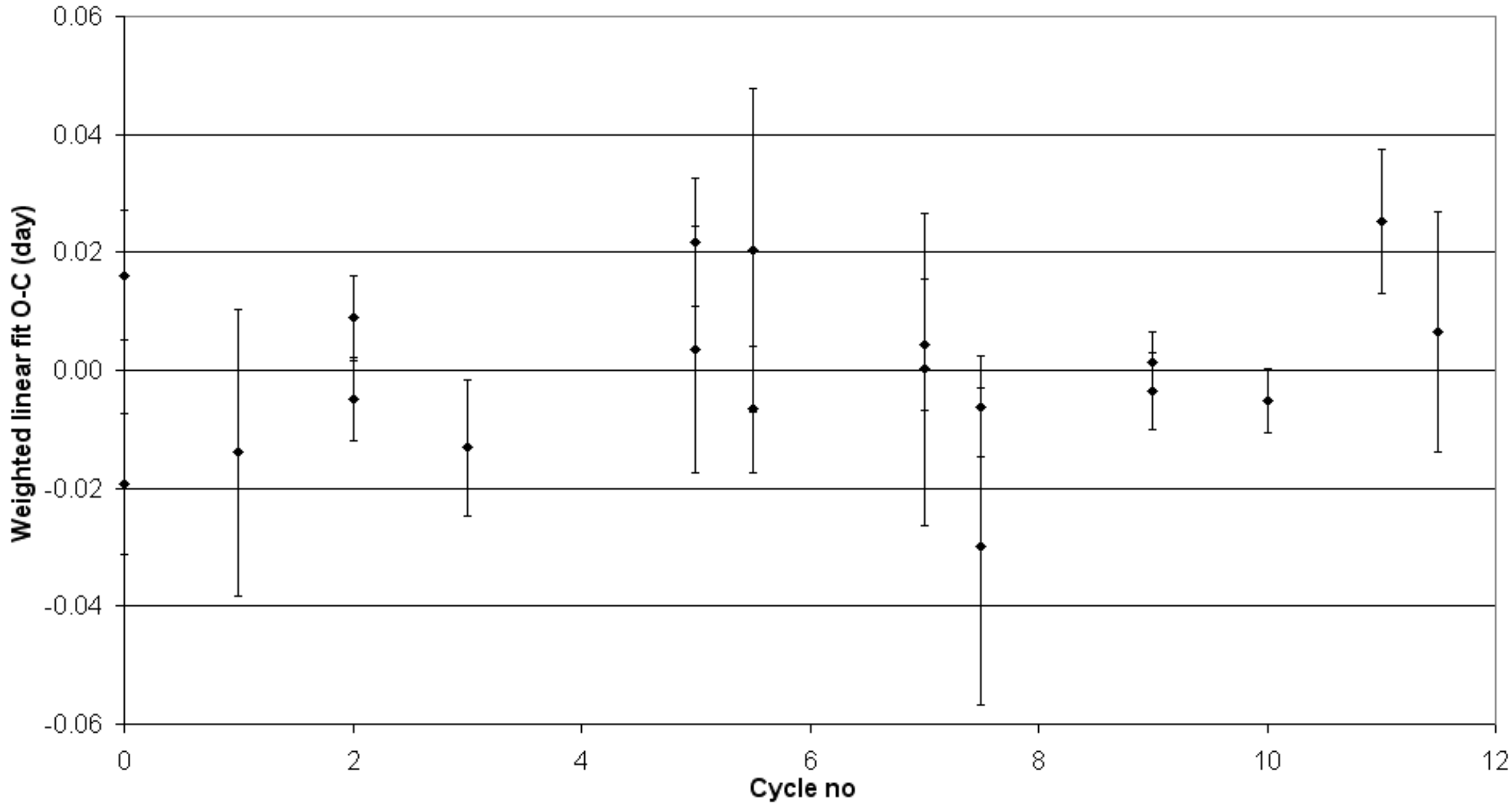

Figure 4. O-C diagram with respect to the linear ephemeris in eqn (1).

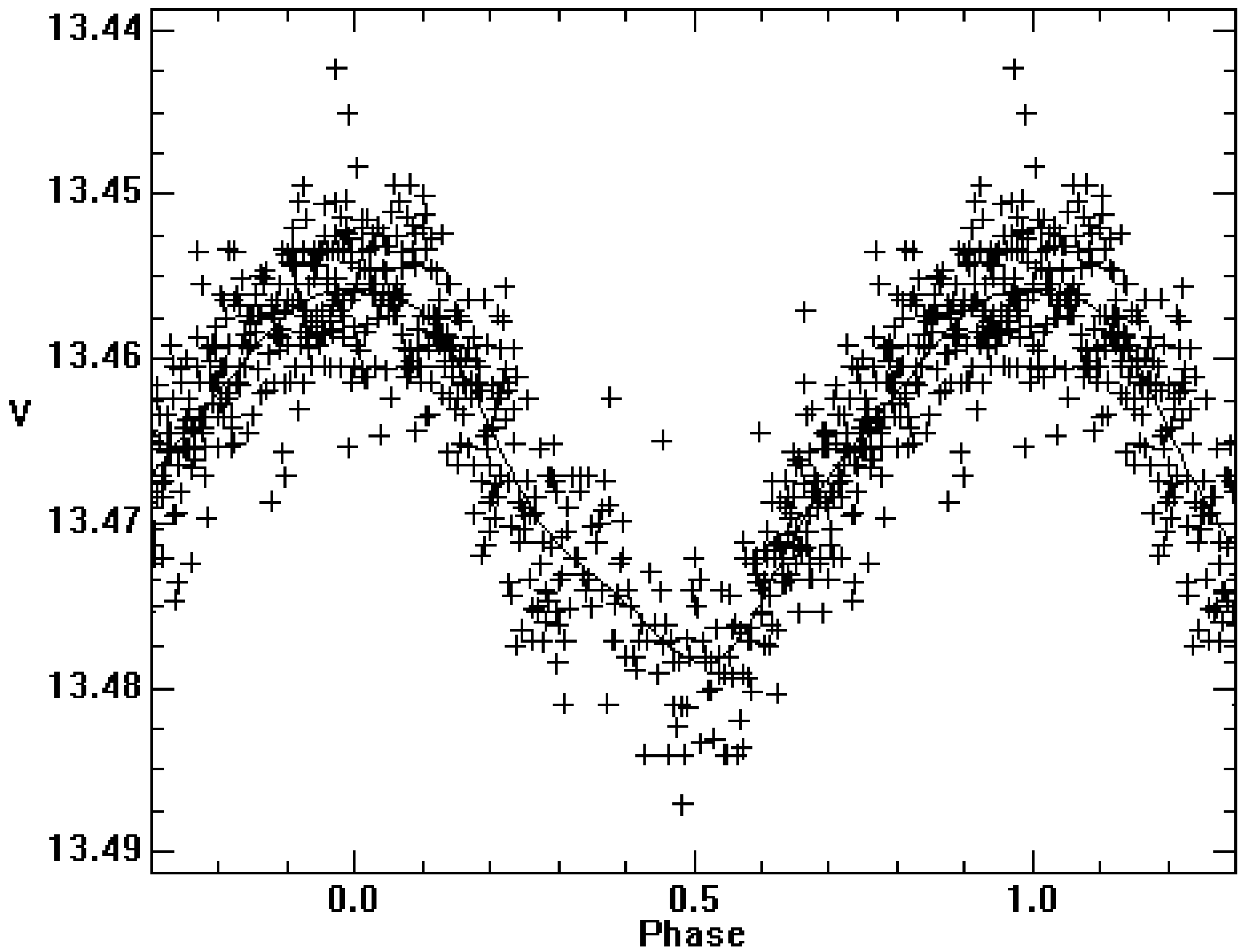

Figure 5. Averaged phase diagram for a period of 0.43815 d using eqn (2).

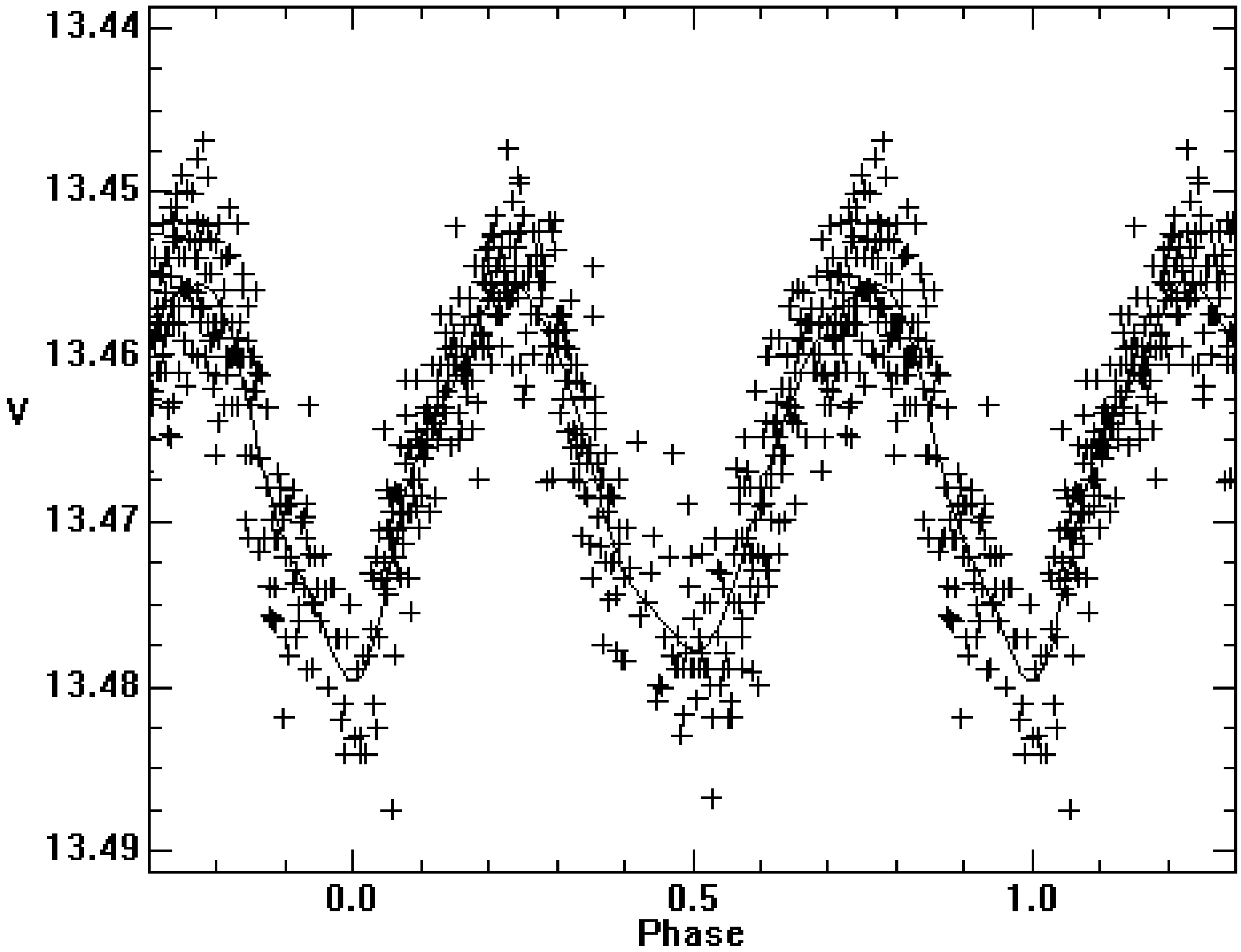

Figure 6. Averaged phase diagram for a period of 0.87602 d using eqn (3).